\documentclass[amssymb,nofootinbib]{revtex4}
\usepackage{amstext,amsmath,amssymb,amsfonts,bbm}
\usepackage[utf8]{inputenc}
\usepackage{fancyhdr}
\usepackage[dvips]{graphicx}
\usepackage{epsfig}
\usepackage{hyperref}

\def\bra{\langle} \def\ket{\rangle}
\def\f{\frac}
\newcommand{\SU}{\mathrm{SU}}

\newcommand{\SO}{\mathrm{SO}}

\newcommand{\U}{\mathrm{U}}
\newcommand{\lalg}[1]{\mathfrak{#1}}
\newcommand{\su}{\lalg{su}} 
\renewcommand{\u}{\lalg{u}} 
 
 \newcommand{\spin}{\lalg{spin}}

\def\mone{^{-1}}
\def\tl{\widetilde}
\def\pp{\partial}
 \def\arr{\rightarrow}
\def\eps{\epsilon}

\newcommand{\tr}{\mathrm{tr}}
\newcommand{\Tr}{\mathrm{Tr}}

\def\N{{\mathbbm N}}
\def\Z{{\mathbbm Z}}
\def\R{{\mathbbm R}}
\def\id{\mathrm{id}}

\def\beq{\begin{equation}}
\def\ee{\end{equation}}
\def\bes{\begin{eqnarray}}
\def\ees{\end{eqnarray}}

\def\l({\left(}
\def\r){\right)}
\def\lv{\lvert}
\def\rv{\rvert}

\begin{document}
\title{{\large Spin foam models for quantum gravity from lattice path integrals}}
\author{{\bf Valentin Bonzom}\email{valentin.bonzom@ens-lyon.fr}}
\affiliation{Centre de Physique Th\'eorique, CNRS-UMR 6207, Luminy Case 907, 13288 Marseille, France EU}
\affiliation{Laboratoire de Physique, ENS Lyon, CNRS-UMR 5672, 46 All\'ee d'Italie, 69007 Lyon, France EU}

\begin{abstract}

Spin foam models for quantum gravity are derived from lattice path integrals. The setting involves variables from both lattice BF theory and Regge calculus. The action consists in a Regge action, which depends on areas, dihedral angles and includes the Immirzi parameter. In addition, a measure is inserted to ensure a consistent gluing of simplices, so that the amplitude is dominated by configurations which satisfy the parallel transport relations. We explicitly compute the path integral as a sum over spin foams for a generic measure. The Freidel-Krasnov and Engle-Pereira-Rovelli models correspond to a special choice of gluing. In this case, the equations of motion describe genuine geometries, where the constraints of area-angle Regge calculus are satisfied. Furthermore, the Immirzi parameter drops out of the on-shell action, and stationarity with respect to area variations requires spacetime geometry to be flat.

\end{abstract}
\maketitle

\section*{Introduction}

Spin foam models \cite{baez def SF} form a promising approach to background independent, non-perturbative quantization and have been mainly developped as the structure emerging when writing transition amplitudes between spin network states in loop quantum gravity \cite{carlo book}. It is also quite intuitive to think of them as some functional integrals discretized on a triangulation of spacetime \cite{reisenberger worldsheet}, \cite{baez BF}.

To challenge the issue of writing a spin foam model for quantum gravity, one usually considers Plebanski's approach \cite{plebanski}, \cite{simplicity}. Riemannian gravity is written as a Spin(4) BF theory, where the field $B$ is constrained to come from a co-tetrad 1-form $e$ by the so-called simplicity constraints: $B=\star (e\wedge e)$ ($\star$ being the Hodge dual on Spin(4)). On the lattice, the field $B$ is discretized on the triangles of the triangulation, in a given frame for each of them, and these quantities are called bivectors. The simplicity constraints are then a set of equations involving the bivectors within each tetrahedron.

The approach which has been followed to derive the Barrett-Crane (BC \cite{BC}) model and the Engle-Pereira-Rovelli (EPR \cite{epr}) model is to impose the simplicity constraints after quantization of the unreduced phase space. Geometric quantization allows to define states, which correspond to Spin(4) boundary spin networks, and operators for each tetrahedron \cite{quantum tet}. It was suggested in \cite{etera cs} to use coherent states instead of of the standard magnetic numbers to describe tetrahedral states. This is precisely the method used in \cite{fk} by Freidel and Krasnov (FK), and in \cite{etera new sf} by Livine and Speziale to define the FK model: coherent states enable to keep track of the directions of the bivectors while quantizing and thus provide a better control of the constraints. 

However, these techniques are quite specific, and we would like to have a functional integral representation for several reasons. This is indeed the standard way to proceed in quantum field theory: physical quantities can be computed from a few ingredients, namely an action and a measure. Typically, in lattice Yang-Mills theories, which are also discretized gauge theories, one starts with the gauge invariant Wilson action and the $\SU(N)$ Haar measure \cite{drouffe zuber}. Thus, the path integral is a priviledged tool to investigate physics. In fact, new results have already been obtained from a path integral representation of the new models (FK and EPR) by Freidel and Conrady \cite{conrady1}.

However, and this is our second motivation, the work of \cite{conrady1} has not completely unraveled the geometric content of the new models. Indeed, the action they proposed seems quite unnatural and rather specific to the new models. They also work from the beginning with quantized areas while we would like to start instead with completely classical variables. In a second paper \cite{conrady2}, the same authors studied the stationary points of their action for large areas and showed that the semi-classical behaviour corresponds to genuine discrete geometries with a Regge action \cite{regge}. Here, we adopt a reverse point of view, i.e. the definition of the discretized functional integral from constrained BF theory should itself have an interpretation {\it \`a la Regge}, since both are basically lattice gravity.

We show in the present paper that a large class of spin foam models can be derived within a framework combining both lattice BF theory and Regge calculus with classical areas and dihedral angles, for any value of the Immirzi parameter $\gamma$. This gives a picture of the new models (more precisely, the FK model, and the EPR model when the Immirzi parameter is smaller than unity) as corresponding to a special choice of gluing between simplices, as we will explain. Furthermore, in contrast with the semi-classical analyses of \cite{conrady2} and \cite{barrett fairbairn}, we look at the full set of the equations of motion of the FK model, for non-degenerate configurations, and in particular to those obtained by varying the action with respect to areas. The effect of integrating areas in the path integral is a key point to understand the quantum implementation of Einstein (or Regge) equations, which was not treated in \cite{conrady2}, \cite{barrett fairbairn}. Describing the on-shell configurations is obviously a first step. We show that the equations of motion describe Regge geometries, as expected, and in addition, that stationarity with respect to area variations leads to flat spacetimes.

The action and the path integral we proposed are based on three ingredients. First, the simplicity contraints enable to recover local degrees of freedom from BF theory. They have received much attention from the early days of the BC model. The original references for the new models have focused on them, and have provided them with a deep understanding. We use in particular the formulation given in \cite{fk}. The second key point is a Regge action which takes into account the Immirzi parameter. It is a function of areas, (4d) dihedral angles and some non-geometric angles whose presence is due to $\gamma$. This is the only explicit dependence of our path integral both on $\gamma$ which drops out on-shell and on the areas, which is the very reason why stationary points of the complete action correspond to flat spacetimes.

Finally, we need to take care of the parallel transport of bivectors, in order to ensure a consistent gluing between simplices. These relations have to be inserted into the path integral through a weighting (gluing) function which concentrates the amplitude around the correct rules of parallel transport. The new models are specified by a precise choice of such a weight, which is intimately related to the action of Freidel and Conrady \cite{conrady1} and allows to get the directions of the bivectors into coherent states. The need to pay attention to parallel transport has been pointed out in \cite{lagrangian BC}. It was shown there that the BC model correctly implements simplicity, but the equations of motion for the proposed action fail to reproduce consistent parallel transport of bivectors. Still, it turns out that parallel transport together with simplicity make sure that the dihedral angles, which appear in the Regge action, are functions, on-shell, of the bivectors or equivalently of the 3d angles. The constraints defining area-angle Regge calculus in \cite{area-angleRC} are thus satisfied. However, these complicated relations are very simply encoded in the constraints, through the group structure of Plebanski gauge theory, by formulating cross-simplicity and the rules for parallel transport with only group variables in a covariant way \cite{1}.

The simplicity constraints and the Regge action which we use are quite unambiguous, in contrast with the choice of the gluing function. This gives a deep physical meaning to a large class of spin foam models, since those geometric relations do not depend on the gluing function\footnotemark. Our framework allows to control the the way parallel transport is implemented: we consider the same path integral with an arbitrary gluing function and the corersponding spin foam model. The formal structure is similar, including the fusion coefficients. It is not surprising since the latter have been shown to be natural objects when dealing with cross-simplicity. But the sum over spin foam is differently weighted and the boundary data slightly differ.

\footnotetext{Provided the gluing function is concentrated around the identity (so that it really glue simplices, only the closure relation depend on its precise form. This point is discuss in details in the article.}



\section{The spin foam models of interest}

Spin foam models are standardly built using a triangulation of spacetime. Then, they can be seen as living on the dual 2-complex to the triangulation. In that dual picture, triangles are dual to faces, both denoted $f$, tetrahedra to edges, both denoted $t$ and 4-simplices to vertices denoted $v$. The boundary of a dual face is made of the edges and vertices respectively dual to the tetrahedra and 4-simplices sharing $f$. The orientations of tetrahedra and triangles induce orientations for dual edges and dual faces. A spin foam model assigns amplitudes to triangles, tetrahedra and 4-simplices, which are functions of a coloring of the 2-complex. For lattice gauge theories, a coloring is a labelling of simplices with data coming from the representation theory of the structure group. One can then write transition amplitudes as sums over colorings, allocating to simplices the corresponding amplitudes \cite{baez def SF}. We will as usual mainly focus on the partition function $Z$:
\beq
Z = \sum_{\{\mathrm{colorings\ } c\}} \prod_f W_f(c)\ \prod_t W_t(c)\ \prod_v W_v(c)
\ee
Physically, one obviously expects to formulate an interpretion of the coloring data in terms of eigenvalues of some relevant operators \cite{carlo 3d}, \cite{baez BF}. This can be made more easily inthe semi-classical regime \cite{graviton}, \cite{conrady2}, \cite{barrett fairbairn}.

The quantity of interest is usually the 4-simplex amplitude $W_v$. For $\SU(2)$ BF theory, a coloring is an assignment of irreducible representations to triangles and intertwiners between the four triangle representations of tetrahedra. The 4-simplex amplitude consists in a $\SU(2)$ 15j-symbol built with these data \cite{ooguri}, which is called the boundary spin network (since it lives on the boundary of each 4-simplex). To see that, let us review some basic facts about the topological $\SU(2)$ BF theory \cite{baez BF}. We consider a connection $A$, which will be locally seen, as usual, as a 1-form taking values in the Lie algebra $\su(2)$ ($A^i$, where $i=1,2,3$ are 3d Euclidean indices). The action of the topological quantum field theory called BF is built with a $\su(2)$-valued 2-form field $B$, transforming under the adjoint representation of $G$:
\beq
S_{\mathrm{BF}} = \int_M \Tr\Bigl(B\wedge F(A)\Bigr)
\ee
where $F(A)=dA+\f{1}{2}[A,A]$ is the curvature of $A$, and $M$ denotes spacetime. The action is gauge invariant and the equations of motion are:
\begin{align}
d_A B &= 0 \label{daB}\\
F(A) &= 0 \label{flatness}
\end{align}
where $d_A = d + [A,\cdot]$ is the covariant derivative. Thus $B$ can be seen as a Lagrange multiplier imposing $A$ to be flat. Moreover, the field $B$ can be completely gauged away thanks to an additional symmetry which is due to the Bianchi identity: $B' = B + d_A \phi$ for any $\lalg{g}$-valued 1-form $\phi$, while $A$ is unchanged. The theory has thus no local degrees of freedom.

To derive a spin foam model, we first discretize the configuration variables on the triangulation. Keeping in mind that we deal with a gauge theory, we follow an idea of Regge which consists in concentrating curvature around triangles. Tetrahedra and 4-simplices are flat and equipped with local frames. The connection is discretized like in usual lattice gauge theory: some $\SU(2)$ elements $g_{vt}$ allow for parallel transport between those local frames. In the dual skeleton picture, the two ends of a dual edge $t$ correspond to the two 4-simplices sharing the tetrahedron $t$. Thus, a dual edge is attached two group elements $g_{vt}$, one for each end. The curvature around a triangle $f$ is thus measured by the oriented product of these group elements all along the boundary of the dual face, starting at a base point (reference frame) $v^\star$, $g_f(v^\star) = 
\prod_{t\subset \pp f} g_{vt} g_{v't}\mone$ if $v$, $v'$ are source and target vertices for each dual edge $t$. The flatness imposed by the e.o.m. then reads: $g_f(v)=\id$\footnotemark for each $f$. In the continuum, the field $B$ appears a Lagrange multiplier imposing this specific condition. We consider the equivalent situation on the lattice:
\begin{align}
Z_{\mathrm{BF}} &= \int \prod_{(t,v)} dg_{vt} \Bigg[\prod_f db_f(v^\star)\ e^{i\tr(b_f(v^\star)g_f(v^\star))}\Bigg] \label{sf bf1}\\
&= \int \prod_{(t,v)} dg_{vt}\ \prod_f \delta\bigl(g_f(t)\bigr) \\
&= \sum_{\{j_f\}}\sum_{\{i_t\}}\ \prod_f \bigl(2j_f+1\bigr)\ \prod_t \bigl(2i_t+1\bigr)\ \prod_v \mathrm{15j}\bigl(j_f,i_t\bigr) \label{sf bf2}
\end{align}
where $dg$ is the $\SU(2)$ Haar measure. To obtain this result, we have expanded the $\SU(2)$ delta functions over the characters of spin $j\in\f{\N}{2}$: $\delta(g) = \sum_{j} (2j+1)\chi_j(g)$. If $g=e^{i\phi\hat{n}\cdot\vec{\sigma}}$ for $\phi\in[0,2\pi)$, then: $\chi_j(g) = \f{\sin (2j+1)\phi}{\sin\phi}$. A group element $g_{vt}$ appears four times, in four different representations $j_1,j_2,j_3,j_4$: once in each holonomy around each of the four triangles of the tetrahedron $t$. The integral of the product of four matrix elements can then be written as the identity over the invariant space Inv$(j_1,j_2,j_3,j_4)$, i.e. as a sum over orthogonal intertwiners between these representations.

\footnotetext{ \label{foot1} In fact, due to the use of group elements in the discrete action, and not Lie algebra elements, the action, given below, only catches the projection of $g_f$ onto the Pauli matrices so that only the sine of the class angle of $g_f$ is restricted to be zero. The class angle can thus be 0 or $2\pi$, i.e. $g_f = \pm\id$.}

To say it roughly, spin foam models for quantum gravity are usually built up starting from \eqref{sf bf2}. We would like instead to give them a definition analogous to \eqref{sf bf1}, including constraints on the variables $b_f$. The underlying continuous theory is supposed to be given by the Holst action:
\beq \label{holst action}
S_\gamma = \int \Bigl(\star(e\wedge e)+\f 1\gamma(e\wedge e)\Bigr)^{IJ}\,F_{IJ},
\ee
the first (second) term corresponding to the geometric (non-geometric) sector. The simplicity constraints, which turn the Spin(4) BF action into \eqref{holst action}, are solved by both sectors \cite{simplicity}. However, for a finite Immirzi parameter, one expects both sectors to be equivalent up to a change $\gamma\arr \gamma\mone$, as noticed in \cite{epr}. It is indeed equivalent to start from the non-geometric sector, and introduce the geometric term with the coupling constant $\gamma$, up to a global scaling:
\beq
S_\gamma = \f 1\gamma \Bigl[\int \Bigl((e\wedge e) + \gamma\,\star(e\wedge e)\Bigr)^{IJ}\,F_{IJ}\Bigr].
\ee

The spin foam models introduced by Freidel and Krasnov \cite{fk} have some so-called Spin(4) projected spin networks as boundary states. They are built as follows. We will extensively use the fact that the group Spin(4) is the product $\SU(2)\times\SU(2)$ (the self-dual and anti-self-dual subgroups). That obviously holds for the irreducible representations, thus labelled by two spins, i.e. half-integers $(j^+,j^-)$. We also restrict attention to the case of a positive Immirzi parameter $\gamma\geq 0$ (the case $\gamma\leq0$ corresponds to exchanging the self-dual and anti-self-dual sectors). One first considers the $\{15j\}$-symbol for Spin(4) which is the product of two $\{15j\}$-symbols for $\SU(2)$, independently colored by triangle spins $(j^+_f,j^-_f)$ and intertwining spins $(i^+_t,i^-_t)$. Constraints are introduced on the representations coloring triangles:
\beq \label{quantum diag}
j^\pm_f=\lv\gamma_\pm\rv j_f
\ee
where $j_f\in\f{\N}{2}$, and $\gamma_\pm$ are the integers determined by the Immirzi parameter and some prescriptions which are discussed in the end of the section. $j_f$ is naturally interpreted as the quantum area of the triangle $f$, and $j^\pm_f$ as quantized fluxes for the norm of the self-dual and anti-self-dual parts of the two-form $(e\wedge e) + \gamma\,\star(e\wedge e)$ (when the latter is discretized on the triangle $f$).

\begin{figure} \begin{center}
\includegraphics[width=7cm]{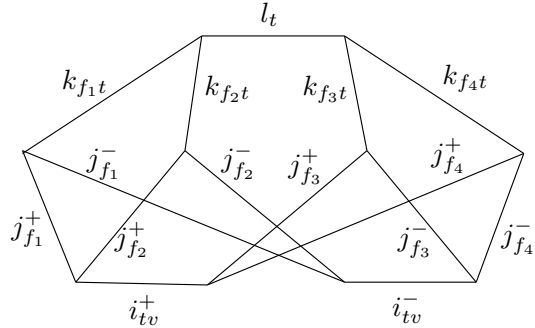}
\caption{ \label{fusion} The evaluation of this spin network is the fusion coefficient. Each spin $l_t$, $i^+_{tv}$ and $i^-_{tv}$ intertwines four representations labelling the four faces of the tetrahedron $t$. To define it unambiguously, edges and vertices need to be oriented. The orientations have to be consistently chosen when summing the fusion coefficients with $\{15j\}$-symbols and other fusion coefficients. }
\end{center}
\end{figure}

The second key ingredients are the so-called fusion coefficients $f_{i^+i^-}^l$, originally appearing in the EPR model \cite{epr}, which projects the Spin(4) intertwiners $(i^+_t,i^-_t)$ onto $\SU(2)$ intertwiners $l_t$. One obviously needs to first intertwine the self-dual with the anti-self-dual representations: $j^+_f\otimes j^-_f\arr k_{ft}$, where $k_{ft}$ is a spin coloring the triangle $f$ independently for each tetrahedron $t$. Then, the spins of the four faces meeting at $t$ are intertwined through $l_t$. This corresponds to the evaluation of the graph depicted in figure \ref{fusion}. Finally, the vertex amplitude is:
\beq \label{new vertex}
W_v(j^+_f,j^-_f,k_{ft},l_t) = \sum_{\{i^+_t,i^-_t\}} \mathrm{15j}\bigl(j^+_f,j^-_f;i^+_t,i^-_t\bigr)\, \prod_{t\subset v} d_{i^+_t}d_{i^-_t}\,f_{i^+_t i^-_t}^{l_t}\bigl(j^+_f,j^-_f, k_{ft}\bigr)
\ee

Equipped with this 4-simplex amplitude, one cannot however define a spin foam model: we need to describe how these basic amplitudes are summed, i.e. give amplitudes to tetrahedra and triangles. In fact, the choice of the measure for tetrahedra distinguishes between the different spin foam models of interest, and depends on $\gamma> 1$ or $\gamma< 1$.

For $\gamma> 1$, the FK model allows for all representations $k_{ft}$, summing over them with a weight:
\beq \label{FK weight}
W_t^{FK,\gamma>1} = d_{l_t}\,\prod_{f\subset t}d_{k_{ft}}\Bigg[\begin{pmatrix} j^+_f &j^-_f &k_{ft} \\ j^+_f &-j^-_f &j^-_f-j^+_f \end{pmatrix}\Bigg]^2 
\ee
As shown in \cite{conrady1}, the $3jm$-coefficient appearing in \eqref{FK weight} is peaked around the value $k=j^+-j^-$. It turns out that the choice $k=j^+-j^-$ is precisely the prescription defining the EPR model for $\gamma>1$:
\beq \label{EPR weight}
W_t^{EPR,\gamma>1} = d_{l_t}\,\prod_{f\subset t} \delta_{k_{ft},j^+_f-j^-_f}
\ee
For $\gamma<1$, the FK and EPR models both select the highest spin in the decomposition of $j^+_f\otimes j^-_f$:
\beq \label{gamma<1}
W_t^{\gamma<1} = d_{l_t} \prod_{f\subset t} \delta_{k_{ft},j^+_f+j^-_f}.
\ee
When $\gamma$ is zero, both models reduce to the originally proposed EPR model. However, for $\gamma=\infty$, the EPR model only selects the mode $k_{ft}=0$, as in the Barrett-Crane model, which is different from the original FK model.

Let us briefly discuss the choice of the integers $\gamma_\pm$. The naive choice comes from identifying the coefficients in the self-dual/anti-self-dual decomposition of the Holst action:
\beq \label{naive presc}
\gamma_\pm = \f{1\pm\gamma}{\gamma} \qquad \qquad\qquad \text{Naive prescription}
\ee
The case $\gamma=0$ would however not be well defined in the FK model. Thus, the FK prescription rather considers the ratio of $\gamma_+$ and $\gamma_-$. They are the integers with the smallest absolute values satisfying $\gamma^+>0$ and:
\beq \label{fk presc}
\f{\gamma_+}{\gamma_-} = \f{\gamma+1}{1-\gamma} \qquad \qquad\qquad \text{FK model}
\ee
In the EPR model, $\gamma_\pm$ are defined by:
\beq \label{epr presc}
\gamma_\pm = \f{\lv 1\pm\gamma\rv}2 \qquad \qquad\qquad \text{EPR model}
\ee
Notice that in the EPR model, it is always true that $k_{ft}=j_f$, due to \eqref{epr presc}. In the presently proposed, Lagrangian approach, these parameters only appear in the action and any of the previous definitions can be chosen.

The face amplitude $W_f$ is ambiguous in the new models. It is a measure for the quantized areas $j_f$, and should be related to the measure on classical areas. If the simplicity constraints can be solved to give the structure of the tetrahedron amplitude, they do not provide information about how triangles should be weighted. Similarly, the normalisation of $W_t$ is unclear.




\section{BF lattice gauge theory and the simplicity constraints} \label{sec:simplicity}

The framework of our discrete path integral is precisely that introduced and described in \cite{1}. Let us sum up the main features. The idea is to discretize Spin(4) BF theory on a trianguation of spacetime, and then introduce within the partition function the simplicity constraints, or rather some of them, at the discrete level. So we first discretize BF theory as a gauge theory. In the spirit of Regge calculus, we also concentrate curvature around triangles. Consequently, local frames (for the internal Spin(4) symmetry) are assigned to flat simplices, i.e. tetrahedra and 4-simplices. Parallel transport between the frames of different simplices are performed using Spin(4) group elements, $G_{vt}=(g_{+vt},g_{-vt})$. In the dual 2-skeleton, edges and points are respectively dual to tetrahedra and 4-simplices. Thus an element $G_{vt}$ is assigned to each half dual edge.

The field $B$ is discretized on triangles of the triangulation, as a bivector denoted $B_f^{IJ}$ ($I,J=0,1,2,3$), i.e. an element of the algebra $\spin(4)$. Such bivectors are naturally defined with respect to local frames. Bivectors of a given triangle in different frames, for instance different tetrahedra, are simply related by parallel transport along the boundary of the corresponding dual face, the elements $G_{vt}$ acting by conjugation:
\beq \label{parallel transport}
B_{f}(t) = G_{tt'}\,B_{f}(t')\,G_{tt'}\mone.
\ee
If $t$ and $t'$ belong to the boundary of the same 4-simplex, then: $G_{tt'} = G_{vt}\mone G_{vt'}$. In agreement with parallel transport, a Spin(4) gauge transformation $K$ is a family of group elements $\{K_t,K_v\}$ acting by:
\begin{align} \label{gauge transfo}
&K\,\triangleright G_{vt} = K_v\,G_{vt}\,K_t\mone, \\
&K\,\triangleright B_f(t) = K_t\,B_f(t)\,K_t\mone.
\end{align}

The simplicity constraints standardly consist in the diagonal and cross-simplicity constraints, which will be taken care of just below, and the closure relation, which will be obtained as an equation of motion in section \ref{sec:flatness}. They make sure that each 4-simplex can be described by 4-vectors representing its edges, so that the bivectors are wedge products of these edge vectors \cite{BC}. Then, the parallel transport relations enable to consistently glue the 4-simplices in the whole triangulation \cite{conrady2}. The natural formulation of the simplicity constraints takes place on each tetrahedron. It is thus convenient, as done in the original derivations of the spin foam models of interest, to define independent bivectors $B_{ft}$ for each tetrahedron sharing a given face. As emphasized in \cite{1}, \cite{lagrangian BC} and \cite{daniele}, the relations \eqref{parallel transport} have then to be taken into account within the partition function in order to reglue the tetrahedra along their common triangles. We thus look at defining the partition function the following way:
\beq
Z = \int\prod_{(v,t)} dG_{vt}\,\prod_{(f,t)} dB_{ft}\ e^{iI^{\mathrm{R}}_\gamma}\ \prod\delta\bigl(S\bigr)\ \tl{\delta}\bigl(R\bigr), \notag
\ee
where $dG$ is the Haar measure on Spin(4), $dB$ a measure to be precised and $I^{\mathrm{R}}_\gamma$ a Regge-like action derived from the usual discrete BF action. We expect the Immirzi parameter $\gamma$ to only enter the action $I^{\mathrm{R}}_\gamma$. $S$ symbolically denotes the simplicity constraints, while $R$ stands for the gluing condition \eqref{parallel transport}. A key point is the possibility to take it into account through a function $\tl{\delta}$ which is peaked around the identity but with a finite width, in contrast with a delta function.

Let us write $B_{ft}$ following the decomposition $\spin(4)=\su(2)\oplus\su(2)$, $B_{ft}=b_{+ft}\oplus b_{-ft}$. Diagonal and cross-simplicity states that $b_+$ and $b_-$ are related by:
\beq \label{simplicity}
b_{-ft} +\eps\ N_t\mone\,b_{+ft}\,N_t =0
\ee
for a rotation $N_t\in\SU(2)$. Using the isomorphism between $\SU(2)$ and the 3-sphere $S^3$, $N_t$ can be seen as a unit 4-vector $N_t^I$. Equivalently, it is defined by the action of $(N_t,\id)\in$ Spin(4) on the reference vector $N^{(0)}=(1,0,0,0)$, and is thus orthogonal to the tetrahedron $t$ in the sense:
\begin{align}
&N_{tJ}\ (\star B_{ft})^{IJ}=0 &&\qquad\text{for }\ \eps=1, \notag \\
&N_{tJ}\ B_{ft}^{IJ}=0 &&\qquad\text{for }\ \eps=-1. \label{normal}
\end{align}
The sign $\eps=\pm 1$ is due to the fact that diagonal simplicity only asks for $\tr\, b_+^2 = \tr\, b_-^2$. This ambiguity is precisely that of Plebanski's theory whose constraints are solved, in the continuum, by both $B=\star(e\wedge e)$ which is gravity (the so-called geometric sector) and $B=e\wedge e$ which is a non-geometric sector (its equations of motion correspond to a torsion-free spacetime, without equations on the curvature)\footnotemark. The choice $\eps=1$ ($\eps=-1$) corresponds to the geometric (non-geometric) sector. At this stage, an important remark is in order. For a finite Immirzi parameter, one expects both sectors to be equivalent up to a change $\gamma\arr \gamma\mone$. It will be quite similar in the discrete setting, up to a slight difference related to the choice of the coefficients $\gamma_\pm$: like between the prescriptions \eqref{fk presc} and \eqref{epr presc}, there is a freedom to rescale both $\gamma_+$ and $\gamma_-$.

\footnotetext{There is another ambiguity due to the symmetry $B\arr -B$, which can resolved in the discrete setting by a constraint on tetrahedral volumes, as in \cite{BC}. Such a constraint does not seem to appear in the usual spin foam models, so that we will not deal with that ambiguity.}

Since $b_+$ and $b_-$ have equal squared norm and that in addition \eqref{parallel transport} implies that this norm is independent of the frame, i.e. gauge invariant, we use the following paramerization:
\beq
B_{ft} = \f i2\,A_f\, \Bigl(n_{+ft}\,\sigma_z\, n_{+ft}\mone,\,-\eps\ n_{-ft}\,\sigma_z\, n_{-ft}\mone\Bigr)\label{b parametrisation}
\ee
with $n_{\pm ft}\in\SU(2)$. $\sigma_z$ is the standard Pauli matrix diag$(1,-1)$. $A_f$ stands for the (gauge invariant) norm of each $b_{+ ft}$, which is physically identified as being the area of $f$, up to a sign. The elements $n_{\pm ft}$ map the unit 3-vector $\hat{z}=(0,0,1)$ onto the directions of $b_{\pm ft}$. $n_+$ and $n_-$ are naturally related by \eqref{simplicity}. Notice that these elements are not uniquely defined for a given bivector: writing $n=e^{-\f i2 \alpha\sigma_z}e^{-\f i2 \beta\sigma_y}e^{-\f i2 \gamma\sigma_z}$ with the Euler decomposition, it is clear the Euler angle $\gamma$ does not play any role in the direction encoded in $n\sigma_zn\mone$. When using the variables $\{A_f,n_{\pm ft}\}$ instead of the bivectors, this ambiguity translates into an extra $\U(1)\times\U(1)$ gauge symmetry which gives:
\beq
(K,\Lambda)\,\triangleright n_{\pm ft} = k_{\pm t}\,n_{\pm ft}\,e^{\f{i}{2}\lambda_{ft}^\pm\sigma_z}.
\ee
The simplicity constraints \eqref{simplicity} can be written:
\beq \label{constrained measure}
\int \prod_t dN_t\ \prod_{(f,t)} \int_0^{4\pi} d\psi_{ft}\,\delta\Bigl(n_{-ft}\mone\,N_t\mone\,n_{+ft}\,e^{\f{i}{2}\psi_{ft}\sigma_z}\Bigr),
\ee
where $dN_t$ and $d\psi_{ft}$ are the normalized Haar measures over $\SU(2)$ and $\U(1)$ (between 0 and $4\pi$). The integrations over $\psi_{ft}$ take care of the phase ambiguities which we have just mentioned, while those over $N_t\in\SU(2)$ amount to integrating over all normals $N_t^I$ for every tetrahedron. We can keep track of the sign ambiguity $\eps$ by considering that the self-dual and anti-self-dual areas satisfy: $A_{+f} =-\eps A_{-f}$. In fact, it will be directly taken into account in the action.

It is physically interesting to consider the variables $N_t$ and $\psi_{ft}$ as configuration variables, as well as $A_f$ and $n_{\pm ft}$ instead of the bivectors. Gauge transformations act on them by:
\begin{align} \label{N psi transfo}
&(K,\Lambda)\,\triangleright N_t= k_{+t}\,N_t\,k_{-t}\mone, \\
&(K,\Lambda)\,\triangleright \psi_{ft} = \psi_{ft}-\bigl(\lambda^+_{ft}-\lambda^-_{ft}\bigr),
\end{align}
so that $N_t$ and $\psi_{ft}$ enable to preserve the Spin(4) and $\U(1)\times\U(1)$ covariance while imposing the constraints. Due to their special transformation properties, it is clear that they can be gauge-fixed to: $N_t=\id$ and $\psi_{ft}=0$. In the lattice path integral, it corresponds to using the translation invariance of Haar measures in order to reabsorb them into the holonomies and some other angles (which we discuss just below).


\section{The action and its equations of motion} \label{sec:action}

The action we present implicitly and partially takes into account these simplicity constraints. It is made of two different parts which depend on the previously described variables and some additional angles $\theta^\pm_{fv}$ describing the dihedral angles and the gluing of adjacent tetrahedra, as we will see (notice that the label $(fv)$ is equivalent to specifying $t$ and $t'$):
\beq
I_\gamma = I_\gamma^{\mathrm{R}}(A_f,\theta_{fv}^+,\theta_{fv}^-) + I_{s_+}^{\mathrm{CS}}\bigl(n_{+ft},g_{+vt},\theta^+_{fv}\bigr)  +I_{s_-}^{\mathrm{CS}}\bigl(n_{-ft},g_{-vt},\theta^-_{fv}\bigr).
\ee
where $A_f$ is the area of the triangle $f$, and corresponds to $A_{+f}$ in the above section. The term $I_\gamma^{\mathrm{R}}$ is a compactified Regge action, such as those proposed in \cite{caselle} and \cite{kawamoto}, but which includes in addition the Immirzi parameter:
\beq \label{regge action}
I^{\mathrm{R}}_\gamma = - \sum_f A_f\,\sin\Biggl(\sum_{v\supset f}\f{\gamma_+}{2}\,\theta^+_{fv} + \f{\gamma_-}{2}\,\theta^-_{fv}\Biggr),
\ee
while the other part, $I^{\mathrm{CS}}_s$, aims at concentrating the amplitudes around the parallel transport relations \eqref{parallel transport}. The notation 'CS' is for 'coherent states', since its special form is at the root of the coherent states formulation (of \cite{conrady1} for example). $I^{\mathrm{CS}}_s$ does not have an explicit dependence on the Immirzi parameter, but one has to choose a sign $s_\pm=\mathrm{sign}(\gamma_\pm)$ independently for the self-dual and anti-self-dual actions. $I^{\mathrm{CS}}_s$ involves some discrete Lagrange multipliers $j^\pm_{fv}\in\f \N2$:
\beq \label{FK gluing}
I_s^{\mathrm{CS}}\bigl(n_{ft},g_{vt},\theta_{fv}\bigr) = -2i\sum_{(f,v)} j_{fv}\,\ln\,\tr \Biggl(\f 12 \bigl(\id+s\sigma_z\bigr)\,n_{ft}\mone\,g_{tt'}\,n_{ft'}\,e^{\f{i}{2}\theta_{fv}\sigma_z}\Biggr).
\ee
Notice that the elements $n_{+ft}$ and $n_{-ft}$ should be related by simplicity, following \eqref{constrained measure}. Let us mention that the geometric and non-geometric sectors are related by the change $\gamma_-\arr-\gamma_-$ (see section \ref{sec:regge action}).


\subsection{The parallel transport action} \label{parallel transport action}

The function $I^{\mathrm{CS}}_s$ is specific to the new models, and has non-trivial consequences on the equations of motion as we will see. Notice that it has the same functional form as the action proposed in \cite{conrady1}, with different arguments. Let us first look at the variations with respect to the half-integers $j^\pm_{fv}$. Stationarity requires that the complex logarithm in \eqref{FK gluing} is 0 for each pair $(f,v)$. With the parametrization $g=\cos\phi+i\sin\phi\,\hat{u}\cdot\vec{\sigma}\in\SU(2)$, the corresponding function is:
\beq \label{ln tr}
\ln\,\tr \Bigl(\f 12 \bigl(\id+s\sigma_z\bigr)\,g\Bigr) = \ln \Bigl(\cos\phi + is\sin\phi\,u_z\Bigr),
\ee
where $u_z$ is the projection of $\hat{u}\in S^2$ on the direction $\hat{z}=(0,0,1)$. It vanishes for $\phi=0$ only. The equations of motion are thus:
\beq \label{gluing}
g_{+tt'} = n_{+ft}\,e^{-\f{i}{2}\theta_{fv}^+\sigma_z}\,n_{+ft'}\mone,\qquad\text{and}\qquad g_{-tt'} = n_{-ft}\,e^{-\f{i}{2}\theta_{fv}^-\sigma_z}\,n_{-ft'}\mone
\ee
for adjacent tetrahedra $t,t'$. As shown in \cite{1}, the existence of angles $\theta^\pm_{fv}$ such that the above formulas hold is equivalent to the parallel transport condition \eqref{parallel transport}. One can also interpret \eqref{gluing} as the solutions of \eqref{parallel transport} for the holonomies in terms of bivector directions and additional angles $\theta^\pm_{fv}$. Similarly to the Spin(4) parallel transport, the angles $\theta^\pm_{fv}$ behave as a kind of discrete $\u(1)\oplus\u(1)$ connection which enables to transport phases between adjacent tetrahedra:
\beq
(K,\Lambda)\,\triangleright \theta_{fv}^\pm = \theta_{fv}^\pm +\eps_{tt'}^f\bigl(\lambda_{ft}^\pm - \lambda_{ft'}^\pm\bigr)
\ee
where $\eps_{tt'}^f=\pm1$. Since $t$ and $t'$ are adjacent, the corresponding dual edges share the vertex $v$ along the boundary of the dual face $f$. $\eps_{tt'}^f$ is positive when the path $(t\arr t')$ through $v$ is oriented like $f$, and else negative. The angles $\theta^\pm_{fv}$ transform so that the holonomies $G_{tt'}$ are indeed independent of the local choices of Euler angles $\gamma$.

In fact, one can think of $I_s^{\mathrm{CS}}$ as a measure which concentrates the amplitude on the configurations satisfying the correct rules for parallel transport. Indeed, first consider the function on $\SU(2)$ appearing in $I_s^{\mathrm{CS}}$: $g\mapsto 2j\ln(\tr\f 12(\id\pm\sigma_z)g)$. We can exactly perform the sum over each representation $j^\pm_{fv}$ appearing in the discrete path integral thanks to:
\begin{align} \label{gluing function}
\tl{\delta}_s^{\mathrm{cs}}(g) &=\sum_{j\in\f{\N}{2}}e^{2j\,\ln\,(\tr\,\f 12(\id+s\sigma_z)g)}, \\
&= \sum_{j\in\f{\N}{2}} \bra j,sj\rv\,g\,\lv j,s j\ket, \\
&= \f1{1-\cos\phi- is\,\sin\phi\,u_z},
\end{align}
for $g=\cos\phi+i\sin\phi\,\hat{u}\cdot\vec{\sigma}$. We have used the fact that: $\bra j,\pm j\rv\,g\,\lv j,\pm j\ket = [\tr\,\f 12(\id\pm\sigma_z)g]^{2j} = (\cos\phi\pm i\sin\phi\,u_z)^{2j}$. It is obviously peaked around the identity since it diverges at $\phi=0$. The gluing action is thus equivalent to introducing the function $\tl{\delta}_s^{\mathrm{cs}}$ into the partition function at each dual vertex along each dual face:
\beq \label{gluing measure}
\sum_{\{j^+_{fv},j^-_{fv}\}} e^{iI^{\mathrm{CS}}_{s_+}}\ e^{iI^{\mathrm{CS}}_{s_-}} =
\prod_{(f,v)}\,\tl{\delta}_{s_+}^{\mathrm{cs}}\Bigl(n_{+ft}\mone\,g_{+tt'}\,n_{+ft'}\,e^{\f i2 \theta_{fv}^+\sigma_z}\Bigr)\,\tl{\delta}_{s_-}^{\mathrm{cs}}\Bigl(n_{-ft}\mone\,g_{-tt'}\,n_{-ft'}\,e^{\f i2 \theta_{fv}^-\sigma_z}\Bigr).
\ee
The normalisation of the tetrahedron amplitude can be easily changed by introducing some powers of the dimensions $d_{j^\pm_{fv}}$ into the sums \eqref{gluing function}. Positive powers correspond to derivatives of $\tl{\delta}^{\mathrm{cs}}_s$.

\bigskip

We now describe the consequences of cross-simplicity and parallel transport in terms of dihedral angles and Regge calculus. They are precisely derived in \cite{1}, and describe the dominant non-degenerate configurations of the path integral, even if the function $\tl{\delta}^{\mathrm{cs}}$ is replaced with any function concentrated on the identity. As explained in \cite{1}, the angles $\theta^\pm_{fv}$ introduced above have a strong geometric meaning, and the constraints \eqref{constrained measure} together with the gluing conditions \eqref{gluing} imply non-trivial relations which make the link with geometry and Regge calculus. First notice that the holonomy around a triangle $f$ is:
\beq
G_f(t) = e^{\f{-1}{2A_{+f}}(\theta_f^++\theta_f^-)\star B_{ft}}\ e^{\f{-1}{A_{+f}}\theta_f B_{ft}}
\ee
with $\theta^\pm_f = \sum_{v\supset f}\theta^\pm_{fv}$, and $\theta_f\equiv \f 12(\theta^+_f-\theta^-_f)$. This means in particular that $G_f(t)$ leaves $B_{ft}$ invariant. When the constraints are fully satisfied (i.e. simplicity, parallel transport and the closure relation), we expect, like in Regge calculus, that $G_f$ is generated by $B_f$ only, and not $\star B_f$. This suggests that $\theta_f$ is related to the dihedral angles. Indeed, one can show that the internal 3d dihedral angle $\phi^t_{ff'}$, which is the angle between the triangles $f$ and $f'$ within the tetrahedron $t$, is given by:
\beq \label{3d angle}
\cos \phi_{ff'}^t = -\eps_{ff'}\ \bra 1,0\lv\, n_{+ft}\mone\,n_{+f't}\, \lv 1,0\ket,
\ee
where $\eps_{ff'}=\pm 1$ is the relative orientation of the dual faces $f$ and $f'$, and $\lv 1,0\ket$ is the state of spin 1 and momentum projection 0. In addition, the (4d) internal dihedral angle $\theta_{tt'}$ between two adjacent tetrahedra corresponds to the product of their normal vectors, $-N_t\cdot N_{t'}$, up to parallel transport: 
\begin{align}
\cos\theta_{tt'} &= -N_t(v)\cdot N_{t'}(v) \\
&= -\f{1}{2}\,\tr\ N_t\,g_{-tt'}\,N_{t'}\mone\,g_{+tt'}\mone.
\end{align}
Due to the gluing conditions \eqref{gluing} and simplicity \eqref{constrained measure}, the above quantity satisfies an interesting relation:
\beq \label{4d angles}
N_t\,g_{-tt'}\,N_{t'}\mone\,g_{+tt'}\mone = \exp\Biggl[\Bigl(\eps_{tt'}^f(\theta^+_{fv}-\theta^-_{fv})+\psi_{ft}-\psi_{ft'}\Bigr)\,\f{b_{+ft}}{A_f}\Biggr],
\ee
whose geometric interpretation is quite clear. The left hand side is the comparison between the normal vectors to $t$ and $t'$, in the frame of $t$. Since these two tetrahedra meet at a triangle, their normals are not independent: they are both orthogonal to the shared triangle, or equivalently, both live on the plane orthogonal to $\star B_{ft}$. Once a normal to $f$ is given, via a $\SU(2)$ element $N_{ft}$ like in \eqref{normal}, it is easy to see that this plane is described by elements of the $\U(1)$ subgroup leaving $b_{+ft}$ invariant: $N'_{ft} = N_{ft}\,\exp(\phi\, b_{+ft}/A_f)$ for angles $\phi$. When the rotations are those encoding the normals to tetrahedra, the angle turns out to be the dihedral angle $\theta_{tt'}$ (up to a sign):
\beq \label{4d angles 2}
\cos\theta_{tt'} = -\cos\f{1}{2} \Bigl[\theta_{fv}^+-\theta_{fv}^-+\eps_{tt'}^f \bigl(\psi_{ft}-\psi_{ft'}\bigr)\Bigr]
\ee

We will integrate over the angles $\theta^\pm_{fv}$ and $\psi_{ft}$ in the partition function, as well as over the directions of the bivectors $n_{\pm ft}$. It was noted in \cite{barrett first order} that first order Regge calculus can be defined provided the angles appearing in the Regge action are dihedral angles of some geometry. It is thus important to see that the dihedral angles are here implicitly functions of the bivectors. When the constraints \eqref{constrained measure} and \eqref{gluing} hold, some algebraic manipulations of these constraints indeed lead to the standard and expected relation between 3d and 4d dihedral angles:
\beq \label{relation 3d-4d angles}
\cos\theta_{tt''} = \f{\cos\phi^{t'}_{f_1f_2}-\cos\phi_{f_1 f}^t\,\cos\phi_{ff_2}^{t''}}{\sin\phi_{f_1 f}^t\,\sin\phi_{ff_2}^{t''}}
\ee
for non-degenerate configurations. Here, $t$, $t'$ and $t''$ are three tetrahedra in a 4-simplex $v$, such that $t'$ and $t$ ($t''$) meet along $f_1$ (along $f_2$), while $t$ and $t''$ share $f$. This means that for on-shell configurations, simplicity and the parallel transport conditions make sure that the dihedral angles are exactly functions of the 3d angles. Furthermore, since the left hand side, $\cos\theta_{tt''}$, is independent of the intermediate tetrahedron $t'$, this gives a set of constraints between 3d angles by writing \eqref{relation 3d-4d angles} for different choices of $t'$. These constraints were proposed in \cite{area-angleRC} as a set of constraints (together with the closure relation) to build a Regge calculus using areas and 3d angles as basic variables. Thus, on-shell configurations in our path integral will implicitly account for these non-trivial relations, only through constraints on $\SU(2)$ and $\U(1)$ elements.

\subsection{The Regge action} \label{sec:regge action}

Given the above geometric description, the compactified Regge action is quite natural:
\beq
I_\gamma^{\mathrm{R}} = -\sum_f A_{f}\,\sin \f 14\Biggl(\bigl(\gamma_+-\gamma_-\bigr)\,2\theta_f + \bigl(\gamma_++\gamma_-\bigr)\bigl(\theta^+_f+\theta^-_f\bigr)\Biggr),
\ee
$\theta_f$ being the sum of the dihedral angles around $f$:
\beq
\theta_f = \sum \theta_{tt'} = \f 12(\theta^+_f-\theta^-_f),\qquad \text{for}\quad \theta^\pm_f = \sum_{v\supset f}\theta^\pm_{fv}.
\ee
When the naive choice \eqref{naive presc} is made for $\gamma_\pm$, the coefficients are: $\gamma_+-\gamma_- =2$ and $\gamma_++\gamma_-=2\gamma\mone$. Then, the Immirzi parameter appears in front of the angle $\theta^+_f+\theta^-_f$. A similar situation has been observed in the asymptotics of the FK \cite{conrady2} and EPR \cite{barrett fairbairn} models. When the constraints are fully satisfied, it is: $\theta^+_f+\theta^-_f=2\pi n$ for an integer $n$. Thus, the on-shell action is independent of $\gamma_++\gamma_-$, up to a sign, as soon as the latter quantity is an even integer. Note that this is a sufficient condition given in \cite{conrady2} so that the asymptotics computed there is non-zero. This echoes the fact that in the continuous theory, the equations of motion are independent of the Immirzi parameter.

When the gluing conditions \eqref{gluing} hold and also using cross-simplicity \eqref{constrained measure}, the action $I_\gamma^{\mathrm{R}}$ can be written as a BF-like action:
\beq \label{bf-like action}
I_\gamma^{\mathrm{R}} = -\sum_f \tr\Bigl(b_{+ft}\,g_{+f}^{\gamma_+}(t)\,N_t\,g_{-f}^{\gamma_-}(t)\,N_t\mone\Bigr), \qquad\text{on-shell.}
\ee
For $g=e^{\f i2\phi\hat{u}\cdot\vec{\sigma}}\in\SU(2)$ of angle $\phi$ and direction $\hat{u}$, the group element $g^\alpha$ is defined so as to have the same direction and the angle $\alpha\phi$. $I_\gamma^{\mathrm{R}}$ is still gauge invariant. For each triangle, it needs a tetrahedron of reference to be defined, but due to the geometric consistency ensured by the gluing conditions, it does not depend on this choice. We emphasize that \eqref{bf-like action} is really a discretization of Spin(4) BF theory with the constraints \eqref{simplicity} (for the choice $\eps=1$). Consider a chart where the typical length of edges of $f$ is of order $\varepsilon$. When $\varepsilon$ goes to zero, we can use the expansion: $g_{\pm f}\approx 1+\varepsilon^2 F_{\pm \vert f}$, where $F_{\pm \vert f}$ is the component of the curvature along the directions of the face dual to the triangle, and thus $g_{\pm f}^{\gamma_\pm}\approx 1+\varepsilon^2 \gamma_\pm F_{\pm \vert f}$. For each triangle, $I_\gamma^{\mathrm{R}}$ reduces to: $\varepsilon^2\,\gamma_+\tr\, b_{+ft}\,F_{+\vert f} - \varepsilon^2\,\gamma_-\tr\, b_{-ft}\,F_{-\vert f}$, where we have used: $b_{-ft}=-N_t\mone b_{+ft}N_t$. It is the expected naive continuum limit.

In the continuous theory \eqref{holst action}, one can equivalently consider as a reference the geometric or the non-geometric sector, and relate them by the transformation $\gamma\arr\gamma\mone$. Let us see how this is done in the discrete setting. We start with the naive prescription for $\gamma_+$ and $\gamma_-$, in the geometric sector. These choices only affect the Regge action $I^{\mathrm{R}}$:
\beq
I^{\mathrm{R}}_\gamma = - \sum_f A_f\,\sin\Biggl(\sum_{v\supset f}\f{\gamma_+}{2}\,\theta^+_{fv} + \f{\gamma_-}{2}\,\theta^-_{fv}\Biggr),\qquad \text{for}\quad \gamma_\pm = \gamma\mone\pm 1.
\ee
The sign ambiguity relating the geometric and the non-geometric sectors originally appears in the constraint: $b_{-ft}=\mp N_t\mone\,b_{+ft}\,N_t$. We look for translating the ambiguity into the Regge action. One can use its on-shell form \eqref{bf-like action} and its continuum limit, described above, to observe that it amounts to changing the sign in front of $\gamma_-$, and so the exponent of the anti-self-dual holonomy: $g^{\gamma_-}_f\arr g_f^{-\gamma_-}$.
In the same time, let us transform the Immirzi parameter into $\gamma\mone$. This gives the non-geometric action for $\gamma\mone$:
\beq
I^{\mathrm{R,non-g}}_{\gamma\mone} = - \sum_f A_f\,\sin\Biggl(\sum_{v\supset f}\f{1+\gamma}{2}\,\theta^+_{fv} + \f{1-\gamma}{2}\,\theta^-_{fv}\Biggr),
\ee
which precisely corresponds to the EPR prescription \eqref{epr presc} for $\gamma<1$.

\subsection{Flatness and the closure relation} \label{sec:flatness}

We now look at the remaining equations of motion. An essential feature of the action $I_\gamma$ is that it is linear in the areas, and that in addition they only appear in the Regge action $I_\gamma^{\mathrm{R}}$. As a consequence, stationarity of the action with respect to areas leads to:
\beq \label{area variation}
\sin \f 14\Biggl(\bigl(\gamma_+-\gamma_-\bigr)\,2\theta_f + \bigl(\gamma_++\gamma_-\bigr)\bigl(\theta^+_f+\theta^-_f\bigr)\Biggr) = 0.
\ee
This is formally analogous to the flatness condition \eqref{sf bf1} of BF theory, obtained by varying the action with respect to bivectors. Moreover, the above equation is equivalent to that due to extremizing the on-shell action \eqref{bf-like action} with respect to $b_{+ft}$:
\beq \label{area variation 2}
g_{+f}^{\gamma_+}(t)\,N_t\,g_{-f}^{\gamma_-}(t)\,N_t\mone = \pm\id.
\ee
Next, we vary the action with respect to the angles $\theta^\pm_{fv}$, and also get interesting results:
\beq
\gamma_\pm\,A_f\,\cos \f 14\Biggl(\bigl(\gamma_+-\gamma_-\bigr)\,2\theta_f + \bigl(\gamma_++\gamma_-\bigr)\bigl(\theta^+_f+\theta^-_f\bigr)\Biggr) = 2s_\pm\,j^\pm_{fv},
\ee
for each dual vertex $v$ in the boundary of the dual face $f$. Due to \eqref{area variation}, the cosine is just $\pm 1$. This makes clear the role of $s_\pm=\mathrm{sign}(\gamma_\pm)$. Indeed, when $\gamma$ becomes greater than 1, $\gamma_-$ becomes negative. However, all quantities in the previous equation are positive, expect $s_-$ which also changes sign. Finally, the cosine has to be 1, which corresponds to choosing the sign $+\id$ in \eqref{area variation 2}, and there are two kinds of equations:
\begin{align} \label{preflatness}
&\cos \f 14\Biggl(\bigl(\gamma_+-\gamma_-\bigr)\,2\theta_f + \bigl(\gamma_++\gamma_-\bigr)\bigl(\theta^+_f+\theta^-_f\bigr)\Biggr) = 1, \\
&\text{and}\qquad\qquad \lv\gamma_\pm\rv\, A_f = 2j^\pm_{fv}. \label{solve j}
\end{align}
There are several other consequences. As a matter of fact, due to the use of the special gluing function $I^{\mathrm{CS}}$ involving discrete Lagrange multipliers, areas are quantized and take rational values when $\gamma_\pm$ are taken to be integral. Moreover, the spins $j^\pm_{fv}$ are the same for all 4-simplices sharing $f$, and indeed satisfy a relation of the type of the diagonal simplicity constraint \eqref{quantum diag}, except that $A_f$ is not half-integral in contrast with $j_f$ in \eqref{quantum diag} (but the latter equation will naturally hold at the quantum level). Stationarity with respect to the variables $n_{\pm ft}$ do not bring further information.

To reconstruct a geometry from bivectors, i.e. to assign 4-vectors to edges of the triangulation, we need in addition to the simplicity constraints and to the parallel transport relations another equation which constrains bivectors of each single tetrahedron to satisfy:
\beq
\sum_{f\subset t} \eps_{ft}\,B_{ft} = 0 \label{closure}
\ee
where $\eps_{ft}=\pm1$ according to the relative orientation of the dual face $f$ and the dual edge $t$. This is the so-called closure relation. It is usually taken for granted in the definition of spin foam models as being an equation of motion of (unconstrained) BF theory, obtained by varying the action with respect to holonomies. However, the situation is a bit different here. Indeed, the Regge action $I_\gamma^{\mathrm{R}}$ is a BF-like action only on-shell, while the gluing relations constrain holonomies. In fact, the closure relation nevertheless shows up by extremizing the gluing action $I^{\mathrm{CS}}$ with respect to the group elements $g_{\pm vt}$. We consider variations $\delta g_{+vt}$ for a given half-dual edge, such that $\xi_{+vt} = g_{+vt}\mone\,\delta g_{+vt}$ is a $\su(2)$ algebra element. Also assume without loss of generality that the orientations of the four faces sharing the dual edge $t$ are outgoing at the dual vertex $v$. Then:
\beq \label{closure variation}
\delta I^{\mathrm{CS}}_{s_+} = -2i\sum_{f\subset t} j_{fv}^+ \f{\tr\bigl(\f 12 \bigl(\id+s_+\sigma_z\bigr)\,n_{+ft'}\mone\,g_{+t't}\,\xi_{+vt}\,n_{+ft}\,e^{\f{i}{2}\theta^+_{fv}\sigma_z}\big)}{\tr\bigl(\f 12 \bigl(\id+s_+\sigma_z\bigr)\,n_{+ft'}\mone\,g_{+t't}\,n_{+ft}\,e^{\f{i}{2}\theta^+_{fv}\sigma_z}\bigr)}
\ee
We can use the previous equations of motion to simplify this expression. Thanks to the gluing conditions \eqref{gluing}, the numerator becomes: $\tr(\xi_{+vt}\,n_{+ft}\,\sigma_z\,n_{+ft}\mone)$, and the denominator is simply 1. The Lagrange multipliers $j^+_{fv}$ turn out to be the areas \eqref{solve j}. This gives: $\sum_{f\subset t} b_{+ft} =0$. It is trivial to see that it entails: $\sum_{f\subset t} B_{ft} =0$ using cross-simplicity. Relaxing the assumption on orientations just introduces signs and leads to the closure \eqref{closure}.

The equations of motion thus describe genuine geometric situations, with discretized co-tetrads, like in \cite{conrady2}. It is however essential to notice that, like in BF theory, the closure relation comes from a kind of flatness condition \eqref{area variation}. Indeed, the latter equation is necessary to solve the Lagrange multipliers as $j^\pm_{fv}=\lv\gamma_\pm\rv A_f/2$, which is then used in the variations \eqref{closure variation}. Moreover, we can go further and show that this model corresponds to truly flat configurations. Since the full set of constraints is contained in the equations of motion, we can apply the following result of \cite{conrady2} (see also \cite{barrett fairbairn}): the angle $\theta^+_f+\theta^-_f$ is a multiple of $2\pi$. This comes from the fact that 4-vectors can be consistently assigned to the edges of the triangulation, and that their expressions in different frames are related by parallel transport, up to some signs. As a result and providing that $(\gamma_++\gamma_-)$ is an even integer, the sum of the dihedral angles $\theta_f$ is:
\beq \label{preflatness}
(\theta^+_f+\theta^-_f)=0,2\pi\ \mod (4\pi)\quad\Rightarrow\quad\theta_f = \f{4\pi p_f}{\gamma_+-\gamma_-},
\ee
for $p_f\in\N$. The fact that $\theta_f$ is fixed is in contrast with what we expect from gravity. Moreover, when the coefficients $\gamma_\pm$ are given by $\gamma_\pm=(1\pm\gamma)/\gamma$, $\gamma_+-\gamma_-$ equals 2, and the sum of the dihedral angles is a multiple of $2\pi$:
\beq
\cos\sum_{(t,t')}\theta_{tt'} = 1.
\ee
This means that the spacetime geometry is flat. Notice that it does not hold in the purely non-geometric situation, $\gamma=0$. It is certainly worthwhile comparing this result with the work \cite{conrady2} and see how flat spacetimes also arise there. The authors study the saddle point equations of the path integral representation they proposed in \cite{conrady1}. Their action is functionnally the same as $I^{\mathrm{CS}}$, with different arguments. The authors look at the asymptotic behaviour for large quantized areas $j_f$, and thus do not ask for stationarity with respect to them. Only a maximality condition is required such that the real part of \eqref{ln tr} is zero, instead of the vanishing of the function itself. This condition is satisfied for group elements of the form $g=e^{i\phi\sigma_z}$, in contrast with stationarity which imposes $g=\id$. The corresponding phases (on saddle points) are then interpreted as dihedral angles, while the latter are here directly taken into account in the arguments of the gluing action. Thus, true stationarity with respect to the quantized areas $j_f$ in \cite{conrady1} would also lead to the vanishing of those phases, that is to flat spacetimes too. To say it differently, we expect that the sums over the half-integers $j_f$ project in some sense onto flat spacetimes.


\bigskip

We have given the meaning of a gluing action to the function $I^{\mathrm{CS}}$, which ensures that the rules for parallel transport are correct even in the presence of the simplicity constraints. We may thus think of taking into account the gluing conditions into the path integral through an arbitrary function $\tl{\delta}$, instead of $\tl{\delta}^{\mathrm{cs}}$ \eqref{gluing function}, which has to peak the path integral around the exact parallel transport relations. If the corresponding spin foam models are easy to derive (see \ref{generic gluings}), it should be stressed that the equations of motion depends on the choice of $\tl{\delta}$. In particular, we can shed light on the reason why the closure relation holds in the new models by looking at another choice. Consider simply $\tl{\delta}(g) = \delta(g)$. It is a natural choice once we have understood that the complicated function appearing in \eqref{FK gluing} (which is also that of the action of \cite{conrady1}) simply ensures the correct gluing of adjacent tetrahedra. We are thus interested in the action:
\beq \label{strong}
I_\gamma^{\mathrm{strong}} = I_\gamma^{\mathrm{R}} + \sum_{(f,v)} \tr\Bigl(x_{+fv}\,g_{+tt'}\,n_{+ft'}\,e^{\f{i}{2}\theta^+_{fv}\sigma_z}\,n_{+ft}\mone\Bigr) + \tr\Bigl(x_{-fv}\,g_{-tt'}\,n_{-ft'}\,e^{\f{i}{2}\theta^-_{fv}\sigma_z}\,n_{-ft}\mone\Bigr),
\ee
which strongly imposes the parallel transport relations\footnotemark\ thanks to Lagrange multipliers $x_{\pm fv}\in\su(2)$. As usual, variations with respect to $g_{\pm vt}$ lead to closure relations, here for the multipliers:
\beq \label{fake closure}
\sum_{f\subset t} \eps_{ft}\,x_{\pm fv} =0
\ee
where $v$ is the vertex of the group element $g_{\pm vt}$. One has then to solve the Lagrange multipliers in terms of the bivectors. Similarly to \eqref{solve j}, it is obtained by varying the action with respect to the angles $\theta^\pm_{fv}$. These variations insert a matrix $\sigma_z$ which projects $x_{\pm fv}$ onto the direction of $b_{\pm ft}$:
\beq \label{project multiplier}
\gamma_+\,A_f = \pm\f i2\,\tr\bigl( x_{+fv}\,n_{+ft}\,\sigma_z\,n_{+ft}\mone\bigr),
\ee
and similarly for the anti-self-dual part, stating that the projection on this direction is really the area. However, the components of $x_{\pm fv}$ which are orthogonal to $b_{\pm ft}$ are left free and it does not seem to be possible to eliminate them from the fake closure \eqref{fake closure} (and the equations coming from varying $n_{\pm ft}$ do not seem to help at all). The difference with the function $\tl{\delta}^{\mathrm{cs}}$ relies on an essential ingredient: the operator $\f 12(\id+\sigma_z)$. The identity enables to get the closure relation when varying $g_{\pm vt}$. As for the matrix $\sigma_z$, it is crucial to compensate the insertion of another $\sigma_z$ when varying the angles $\theta^\pm_{fv}$, using $\sigma_z^2=\id$, so that there is in fact no projection in contrast with \eqref{project multiplier}.

Thus, there is in general no geometry described by edge vectors when the gluing is imposed by a generic function, unless the closure relation is imposed by hand. Interestingly, adding the closure relation as a constraint solves two problems at once. It avoids the use of some special gluing functions like $\tl{\delta}^{\mathrm{cs}}$ in order to get the closure. Moreover, areas do not only appear in the Regge action anymore, so that the equations of motion do not imply spacetime flatness, but the expected discrete Ricci flatness \cite{area-angleRC}.

\footnotetext{In fact, the integration over the Lagrange multipliers projects on the delta over $\SO(3)$ which only imposes $g=\pm id$ in terms of $\SU(2)$ variables, as it is the case in spin foams for BF theory (see footnote \ref{foot1}). It plays no role in the present argument. }

\section{Computation of the lattice path integral} \label{computation}

The main result is the following. The path integral over a spacetime triangulation:
\begin{multline} \label{path int}
Z_\gamma\equiv \int\prod_{(t,v)}dG_{vt}\,\prod_{(f,t)}dn_{+ft}\,dn_{-ft}\,d\psi_{ft} \prod_t dN_t\ \prod_{(f,t)}\delta\Bigl(n_{-ft}\mone\,N_t\mone\,n_{+ft}\,e^{\f i2 \psi_{ft}\sigma_z}\Bigr)\ \prod_f dA_f\quad e^{iI_\gamma^{\mathrm{R}}(A_f,\theta^+_{fv},\theta^-_{fv})} \\
\times \prod_{(f,v)}\int d\theta^+_{fv}d\theta^-_{fv} \prod_{(f,v)} \tl{\delta}^{\mathrm{cs}}_{s_+}\Bigl(n_{+ft}\mone\,g_{+tt'}\,n_{+ft'}\,e^{\f i2\theta^+_{fv}\sigma_z}\Bigr)\,\tl{\delta}^{\mathrm{cs}}_{s_-}\Bigl(n_{-ft}\mone\,g_{-tt'}\,n_{-ft'}\,e^{\f i2\theta^-_{fv}\sigma_z}\Bigr),
\end{multline}
can be written as a sum over spin foams corresponding to the FK model for any value of the Immirzi parameter $\gamma$, and to the EPR model when $\lv\gamma\rv<1$:
\beq
Z_\gamma = \sum_{\{j_f,k_{ft},l_t\in\f{\N}{2}\}}\prod_t W_t\bigl(\lv\gamma_+\rv j_f,\lv\gamma_-\rv j_f,k_{ft},l_t\bigr)\quad \prod_v W_v\bigl(\lv\gamma_+\rv j_f,\lv\gamma_-\rv j_f,k_{ft},l_t\bigr),
\ee
where the tetrahedron amplitude $W_t$ is given in equations \eqref{FK weight} and \eqref{gamma<1}, and the 4-simplex weight $W_v$ in equation \eqref{new vertex}. The first delta functions in \eqref{path int} encode cross-simplicity, $I_\gamma^{\mathrm{R}}$ is the $\gamma$-dependent compactified Regge action \eqref{regge action}, and the measure on the second line of \eqref{path int} corresponds to the parallel transport action after the Lagrange multipliers have been integrated out \eqref{gluing measure}, with $s_\pm=\mathrm{sign}\gamma_\pm$. We now give the proof, and then discuss some interesting properties related to triangle orientations.

The key idea in spin foams is to write the partition function as a state sum whose data have some geometric meaning. Practically, one may use the Fourier transform over the involved Lie groups to find such expansions. Thus, we expand the exponential of $i$ times the Regge action $I^{\mathrm{R}}_\gamma$ over $\U(1)$ modes. For each face, we have:
\beq \label{expand regge}
e^{-iA_f\sin\f12(\gamma_+\theta^+_f+\gamma_-\theta^-_f)} = \sum_{m_f\in\f{\Z}{2}} J_{2m_f}(A)\, e^{-im_f(\gamma_+\theta^+_f+\gamma_-\theta^-_f)}.
\ee
This is the Jacobi-Anger expansion, which may be seen as a definition of the Bessel functions $J_m$ of the first kind: $J_m(A)=\int_0^\pi \f{d\theta}{\pi\, i^m}\,\cos(m\theta)\,e^{iA\cos\theta}$. The Bessel functions satisfy some interesting properties \cite{abramo}. Their integral over $\R_+$ is normalized to 1 for $m\in\N$, and important symmetries are: $J_{-m}(A) = J_m(-A) = (-1)^mJ_m(A)$. Since the area only appear in this expansion, one can easily integrate it:
\beq \label{int area}
\int_{\R_+}dA_f\ e^{-iA_f\sin\f12(\gamma_+\theta^+_f+\gamma_-\theta^-_f)} = 1 + \sum_{m_f\in\f{\N*}{2}} \Bigl(e^{-i\gamma_+m_f\theta^+_f}\,e^{i\gamma_-m_f\theta^-_f} + (-1)^{2m_f}\,e^{i\gamma_+m_f\theta^+_f}\,e^{-i\gamma_-m_f\theta^-_f}\Bigr)
\ee
Notice that the conjugated modes to $\theta^+$ and $\theta^-$ are:
\beq \label{diag momenta}
m^\pm_f=\gamma_\pm m_f,
\ee
for $m_f\in\N/2$, so that they satisfy a constraint which has precisely the form of the diagonal simplicity constraints \eqref{quantum diag}. However, since the variables $\theta^\pm_{fv}$
appear in the group elements $e^{\f i2 \theta^\pm_{fv}\sigma_z}$, the momenta $m^\pm_f$ should be interpreted as magnetic numbers rather than spins. It is important to see that this fact is only due to the expansion of the compactified Regge action $I^{\mathrm{R}}_\gamma$, and is independent of the choice of the gluing functions.

Next, we come to the effects of the special gluing action $I^{\mathrm{CS}}$. We use the following expansion, over the matrix elements of highest (or lowest when $\gamma_\pm\leq0$) magnetic numbers (remember that $s_\pm=\mathrm{sign}\gamma_\pm$):
\beq \label{exp gluing action}
\sum_{\{j^+_{fv}\}}e^{iI^{\mathrm{CS}}(n_{+ft},g_{+vt},\theta^+_{fv})} = \prod_{(f,v)}\sum_{j^+_{fv}}\bra j^+_{fv},s_+\,j^+_{fv}\rv n_{+ft}\mone\,g_{+tt'}\,n_{+ft'}\lv j^+_{fv},s_+\,j^+_{fv}\ket\,e^{is_+j^+_{fv}\theta^+_{fv}},
\ee
and similarly for the anti-self-dual part. The sum on the left hand side denotes the sum over the assignments of spins $j^+_{fv}$ to all pairs $(f,v)$. For a single mode $m_f$, the integrals over the angles $\theta^\pm_{fv}$ now give:
\begin{multline} \label{theta int}
\prod_{v\supset f}\int d\theta^+_{fv}d\theta^-_{fv}\  e^{is_+j^+_{fv}\theta^+_{fv}+is_- j^-_{fv}\theta^-_{fv}} \Bigl(e^{-i\gamma_+m_f\theta^+_f+i\gamma_-m_f\theta^-_f} + (-1)^{2m_f}\,e^{i\gamma_+m_f\theta^+_f-i\gamma_-m_f\theta^-_f}\Bigr) \\= \prod_{v\supset f} \delta_{j^+_{fv},\lv\gamma_+\rv m_f}\ \delta_{j^-_{fv},\lv\gamma_-\rv m_f}.
\end{multline}
This fixes the spins of the matrix elements in the right hand side of \eqref{exp gluing action} for each face to be: $j^\pm_{fv}=\lv\gamma_\pm\rv m_f$, so that the sum over the $\U(1)$ modes $m_f$ is now a sum over the same variable but interpreted as a $\SU(2)$ spin. This is precisely what the relation \eqref{diag momenta}, which encodes diagonal simplicity, becomes when using the special gluing. Notice also that the terms with the coefficient $(-1)^{2m_f}$ are irrelevant. We come back to this point at the end of the section.

Gathering these results for a whole triangulation, changing the notation $j_f\equiv m_f$, and including the cross-simplicity constraints within each tetrahedron, we get:
\begin{multline}
Z_\gamma = \sum_{\{j_f\in\f{\N}{2}\}} \int\prod_{(t,v)}dG_{vt}\,\prod_{(f,t)}dn_{+ft}\,dn_{-ft}\,d\psi_{ft} \prod_t dN_t\ \prod_{(f,t)}\delta\Bigl(n_{-ft}\mone\,N_t\mone\,n_{+ft}\,e^{\f i2 \psi_{ft}\sigma_z}\Bigr) \\
\times\prod_{(f,v)}\,\bra j^+_f,s_+ j^+_f\rv n_{+ft}\mone\,g_{+tt'}\,n_{+ft'}\lv j^+_f,s_+ j^+_f\ket\,\bra j^-_f,s_- j^-_f\rv n_{-ft}\mone\,g_{-tt'}\,n_{-ft'}\lv j^-_f,s_- j^-_f\ket,
\end{multline}
where $j^\pm_f$ are defined by: $j^\pm_f=\lv\gamma_\pm\rv j_f$. The elements $N_t$ can be reabsorbed on the right of $g_{-vt}$, and the angles $\psi_{ft}$ disappear when inserting $n_{-ft}=N_t\mone\,n_{+ft}\,e^{\f i2 \psi_{ft}\sigma_z}$. This is not surprising in the view of the gauge transformations \eqref{N psi transfo} ($\psi_{ft}$ could also have been reabsorbed into the angles $\theta^-_{fv}$). We arrive at:
\beq \label{Zgamma}
Z_\gamma = \sum_{\{j_f\in\f{\N}{2}\}} \int\prod_{(t,v)}dG_{vt}\,\prod_{(f,t)}dn_{ft}
\prod_{(f,v)}\,\bra j^+_f,s_+ j^+_f\rv n_{ft}\mone\,g_{+tt'}\,n_{ft'}\lv j^+_f,s_+ j^+_f\ket\ \bra j^-_f,s_- j^-_f\rv n_{ft}\mone\,g_{-tt'}\,n_{ft'}\lv j^-_f,s_- j^-_f\ket.
\ee
To match the standard expressions of the new models, like in \cite{fk}, we introduce the coherent states. For a unit vector $\hat{n}\in S^2$, consider a group element $g(n)\in\SU(2)$ which maps the reference 3-vector $\hat{z}=(0,0,1)$ onto $\hat{n}$. Define now the state: $\lv j,n\ket = g(n)\lv j,j\ket$. There is however a phase ambiguity due to a possible rotation leaving $\hat{z}$ invariant in the definition of $g(n)$. This is precisely the Euler angle $\gamma$, when writing $g=e^{-\f i2 \alpha\sigma_z}e^{-\f i2 \beta\sigma_y}e^{-\f i2 \gamma\sigma_z}$, and the phase ambiguity is precisely that which has been taken care of through the angles $\theta^\pm_{fv}$. We can thus rewrite unambiguously, i.e. for any local choice of these Euler angles, the previous expression of $Z_\gamma$ with $\lv j^\pm_f,\hat{n}_{ft}\ket = n_{ft}\lv j^\pm_f,j^\pm_f\ket$. Let us restrict attention to positive Immirzi parameters so that $s_+=1$. For $\gamma<1$, we have $s_-=1$ and:
\beq
Z_{\gamma<1} = \sum_{\{j_f\in\f{\N}{2}\}} \int\prod_{(t,v)}dG_{vt}\,\prod_{(f,t)}dn_{ft}
\prod_{(f,v)}\,\bra j^+_f,\hat{n}_{ft}\rv\,g_{+vt}\mone\, g_{+vt'}\,\lv j^+_f,\hat{n}_{f't}\ket\ \bra j^-_f,\hat{n}_{ft}\rv\,g_{-vt}\mone\,g_{-vt'}\,\lv j^-_f,\hat{n}_{f't}\ket.
\ee
This is precisely the expression of the FK and EPR models for $\gamma<1$ as given in \cite{conrady1} (remember that they only differ in the definition of $\gamma_\pm$). When $\gamma>1$, $s_-=-1$, and using $\bra j,-j\rv g \lv j,-j\ket=\overline{\bra j,j\rv g \lv j,j\ket}$, $Z_{\gamma>1}$ is:
\beq
Z_{\gamma>1} = \sum_{\{j_f\in\f{\N}{2}\}} \int\prod_{(t,v)}dG_{vt}\,\prod_{(f,t)}dn_{ft}
\prod_{(f,v)}\,\bra j^+_f,\hat{n}_{ft}\rv\,g_{+vt}\mone\, g_{+vt'}\,\lv j^+_f,\hat{n}_{f't}\ket\ \overline{\bra j^-_f,\hat{n}_{ft}\rv\,g_{-vt}\mone\,g_{-vt'}\,\lv j^-_f,\hat{n}_{f't}\ket},
\ee
which is indeed the expected expression.

\bigskip

Let us now look at the behaviour of the path integral under the reversal of the orientation of a triangle. Consider a single face and remember that an orientation is needed to use the function $\tl{\delta}^{\mathrm{CS}}$ for the gluing. Indeed, its argument in \eqref{gluing measure} was: $n_{+ft}\mone\,g_{+tt'}\,n_{+ft'}\,e^{\f i2 \theta_{fv}^+\sigma_z}$, for $t$ and $t'$ adjacent tetrahedra whose corresponding dual edges meet at the dual vertex $v$, and if the orientation of $f$ goes from $t$ to $t'$. So with a slight abuse of notation, we now write the gluing function going from $t'$ to $t$ for each dual vertex of $f$:
\beq
\int \prod_{t\supset f}dn_{ft}
\prod_{v\supset f}\,\bra j^+_{fv},s_+ j^+_{fv}\rv n_{ft'}\mone\,g_{+t't}\,n_{ft}\,\lv j^+_{fv},s_+ j^+_{fv}\ket\ \bra j^-_{fv},s_- j^-_{fv}\rv n_{ft'}\mone\,g_{-t't}\,n_{ft}\,\lv j^-_{fv},s_- j^-_{fv}\ket\ e^{is_+j^+_{fv} \theta^+_{fv}\sigma_z}\,e^{is_-j^-_{fv} \theta^-_{fv}\sigma_z}.
\ee
The integration over the angular variables is exactly like before. The reversal of the face orientation can be translated into a complex conjugation, since: $\bra j,j\lv g\rv j,j\ket=\overline{\bra j,j\lv g\mone\rv j,j\ket}$ and $g_{\pm t't}\mone=g_{\pm tt'}$. Thus, we obviously arrive at the result already shown in \cite{conrady1} that the model is invariant under the change of the orientation of a triangle. Indeed, the complex conjugation can be reabsorbed into a change of variables $n_{ft}\arr n_{ft}\eps$ for $\eps=\bigl(\begin{smallmatrix} 0&-1\\1 &0\end{smallmatrix}\bigr)$:
\begin{multline}
\int \prod_{t\supset f}dn_{ft}
\prod_{v\supset f}\,\overline{\bra j^+_{f},s_+ j^+_{f}\rv n_{ft}\mone\,g_{+tt'}\,n_{ft'}\,\lv j^+_{f},s_+ j^+_{f}\ket}\ \overline{\bra j^-_{f},s_- j^-_{f}\rv n_{ft}\mone\,g_{-tt'}\,n_{ft'}\,\lv j^-_{f},s_- j^-_{f}\ket} \\
= \int \prod_{t\supset f}dn_{ft}
\prod_{v\supset f}\,\bra j^+_{f},s_+ j^+_{f}\rv n_{ft}\mone\,g_{+tt'}\,n_{ft'}\,\lv j^+_{f},s_+ j^+_{f}\ket\ \bra j^-_{f},s_- j^-_{f}\rv n_{ft}\mone\,g_{-tt'}\,n_{ft'}\,\lv j^-_{f},s_- j^-_{f}\ket.
\end{multline}

The $\eps$ matrix is such that: $\eps\,g\,\eps\mone = \overline{g}$ for $g\in\SU(2)$. Moreover, it transforms the Pauli matrix $\sigma_z$ into its opposite, so that: $n_{ft}\eps\,\sigma_z\,\eps\mone n_{ft}\mone =-n_{ft}\sigma_zn_{ft}\mone$. It means that the transformation performed on the previous formula corresponds to an underlying parity transformation on the bivectors $B_{ft}\arr-B_{ft}$. However, the interpretation is certainly more subtle, for the change of an area $A_f\arr -A_f$ does not leave $Z_\gamma$ invariant, but inserts a coefficient $(-1)^{2j_f}$ for each face. It can be seen by simply using the property $J_m(-A)=(-1)^mJ_m(A)$ of Bessel functions. Another interesting way to get this result is to observe that the new minus sign can be reabsorbed into the angles appearing into the Regge action $I^{\mathrm{R}}$. When integrating these angles, the transformation amounts to considering the complex conjugated term of the parenthesis in \eqref{theta int}. This quantity is not real precisely because of the coefficients $(-1)^{2j_f}$. As another consequence, integrating areas over $\R$ instead of $\R_+$ results in a factor $(1+(-1)^{2j_f})$ for each face (and except for $j_f=0$), which selects the integral spins. We notice that this restriction can be counterbalanced through the redefinition: $\gamma_\pm\arr\gamma_\pm/2$.

When looking at the equation \eqref{theta int}, it is clearly equivalent to take the complex conjugation of the parenthesis and to reverse the signs $s_+$ and $s_-$. Suppose now that instead of the complex function $\tl{\delta}^{\mathrm{cs}}$, we take its real part, still to peak the amplitude around the parallel transport relations. This is equivalent to taking both signs for $s_+$ and $s_-$. In fact, it makes it possible to describe all values of the Immirzi parameter (except $\gamma=\pm 1$) with an unique path integral. But again, because of the two kinds of terms in the expansion of $\exp(iI^{\mathrm{R}}_\gamma)$, all terms except $j_f=0$ come with a coefficient $(1+(-1)^{2j_f})$.

We have chosen the simplest measure for areas. This choice has been shown in \cite{1} to reproduce the standard results for (unconstrained) BF theories, with the strong gluing measure \eqref{strong}. Because the path integral is completely explicit, it is possible to use non-tivial measures, or to insert local area observables. One has just to compute the integral of the insertion times a Bessel function. For instance, insertion of the squared area for a triangle gives:
\beq
\int_{\R_+} dA\,A^2\,J_{2m}\bigl(A\bigr) = \bigl(2j+1\bigr)\bigl(2j-1\bigr),\qquad\text{for}\ m\in\N
\ee
Such integrals can be used to compute the spectra of the corresponding observable insertions, or to get non-trivial face amplitudes.

\section{Spin foams for generic gluings} \label{generic gluings}

The present framework enables to keep under control the way parallel transport is implemented at the quantum level. Indeed, instead of the function $\tl{\delta}^{\mathrm{cs}}$, one can use any function on $\SU(2)$ which is concentrated on the identity. As we have shown, the equations of motion may not exhibit the closure relation. But the natural role of $\tl{\delta}^{\mathrm{cs}}$ is rather to glue adjacent tetrahedra, while the closure relation comes from a special choice leading to the new models. Since the gluing is performed and controlled with $\SU(2)$ elements, we consider the following expansion over matrix elements:
\beq
\tl{\delta}(g) = \sum_{j\in\f{\N}{2}}\sum_{m=-j}^j c^j_m\ \bra j,m\rv\,g\,\lv j,m\ket
\ee
We restrict attention to diagonal matrix elements to ensure gauge invariance under the transformation \eqref{N psi transfo} (this restriction would be nevertheless imposed through the integration over the angles $\psi_{ft}$). Let us give some examples:
\begin{align}
&\tl{\delta}(g) =\tl{\delta}^{\mathrm{cs}}_{s=\pm 1} && c^j_m = \delta_{m,sj}, \\
&\tl{\delta}(g) = \delta(g) && c^j_m = 2j+1, \\
&\tl{\delta}(g) = \f{1}{N}e^{\f{\varepsilon}{2}\tr^2(g\,\vec{\sigma})} && c^j_m(\varepsilon) = \f{1}{N'}\Bigl(I_j(\varepsilon) - I_{j+1}(\varepsilon)\Bigr).
\end{align}
In the last line, $\tl{\delta}$ is a Gaussian of (inverse) width $\varepsilon$. $N$ and $N'$ are some normalisations, and $I_j$ is a modified Bessel function, defined by $I_j(z) = i^jJ_j(-iz)$.

For each dual vertex of each face, $\tl{\delta}$ is inserted in the path integral:
\begin{multline}
\prod_{(f,v)}\ \tl{\delta}\bigl(n_{ft}\mone\,g_{+tt'}\,n_{ft'}\,e^{\f i2\theta^+_{fv}\sigma_z}\bigr)\ \tl{\delta}\bigl(n_{ft}\mone\,g_{-tt'}\,n_{ft'}\,e^{\f i2\theta^-_{fv}\sigma_z}\bigr) \\
= \prod_{(f,v)} \sum_{j^+_{fv},j^-_{fv}}\sum_{m^+_{fv},m^-_{fv}} c^{j^+_{fv}}_{m^+_{fv}}\,c^{j^+_{fv}}_{m^+_{fv}}\ \bra j^+_{fv},m^+_{fv}\rv\,n_{ft}\mone\,g_{+tt'}\,n_{ft}\,\lv j^+_{fv},m^+_{fv}\ket\,\bra j^-_{fv},m^-_{fv}\rv\,n_{ft}\mone\,g_{-tt'}\,n_{ft}\,\lv j^-_{fv},m^-_{fv}\ket\ e^{im^+_{fv}\theta^+_{fv}}\,e^{im^-_{fv}\theta^-_{fv}}
\end{multline}
The Regge action is unchanged. Integrating the angles $\theta^\pm_{fv}$ is formally similar to the result of the equation \eqref{theta int}, but the variables $s_\pm j^\pm_{fv}$ are now replaced with magnetic numbers $m^\pm_{fv}$. The latter are thus required to be: $m^\pm_{fv}=\gamma_\pm m_f$, or the opposite, $m_f$ being the half-integers appearing in the expansion of $e^{iI^{\mathrm{R}}_\gamma}$.  The spins $j^\pm_{fv}$ remain free, and are summed over for each wedge, with $j^\pm_{fv}\geq \lv\gamma_\pm m_f\rv$. Let us introduce the following quantity:
\beq \label{R_m}
R_{\gamma_\pm m_f}\bigl(g_{\pm vt},n_{ft}\bigr) = \prod_{v\supset f} \sum_{j^\pm_{fv}\geq\lv\gamma_\pm m_f\rv} c^{j^\pm_{fv}}_{\gamma_\pm m_f}\,\bra j^\pm_{fv},\gamma_\pm j_f\rv n_{ft}\mone\,g_{\pm tt'}\,n_{ft'}\,\lv j^\pm_{fv},\gamma_\pm m_f\ket.
\ee
The partition function becomes:
\begin{multline}
Z_\gamma = \int\prod_{(t,v)}dG_{vt}\,\prod_{(f,t)}dn_{ft} \prod_f \Bigl[R_0\bigl(g_{+ vt},n_{ft}\bigr)\,R_{0}\bigl(g_{- vt},n_{ft}\bigr) \\
+\sum_{m_f\in\f{\N^\star}{2}} \Bigl(R_{\gamma_+ m_f}\bigl(g_{+ vt},n_{ft}\bigr)\,R_{\gamma_- m_f}\bigl(g_{- vt},n_{ft}\bigr) + (-1)^{2m_f}R_{-\gamma_+ m_f}\bigl(g_{+ vt},n_{ft}\bigr)\,R_{-\gamma_- m_f}\bigl(g_{- vt},n_{ft}\bigr)\Bigr)\Bigr]
\end{multline}
which reduces to \eqref{Zgamma} when $c^j_m=\delta_{sj,m}$. We now suppose that $c^j_{-m}=c^j_m$ (as it is the case when $\tl{\delta}=\tl{\delta}^{\mathrm{cs}}_{s=1}+\tl{\delta}^{\mathrm{cs}}_{s=-1}$). Using the transformation $n_{ft}\arr n_{ft}\eps$, one can show that the terms with the coefficient $(-1)^{2m_f}$ are the same than those without it. Thus, only integral $m_f$ contribute. Since $m_f$ appears as a magnetic number in \eqref{R_m}, it is natural to rewrite $Z_\gamma$ as:
\beq
Z_\gamma = \sum_{\{m_f\in\Z\}} \int\prod_{(t,v)}dG_{vt}\,\prod_{(f,t)}dn_{ft} \prod_f R_{\gamma_+ m_f}\bigl(g_{+ vt},n_{ft}\bigr)\,R_{\gamma_- m_f}\bigl(g_{- vt},n_{ft}\bigr).
\ee

The spin foam representation comes from integrating $G_{vt}$ and $n_{ft}$ in the previous expression. The boundary data of a vertex are similar to those of the new models, except that $m_f\in\Z$ is now a magnetic number. We define $m^+_f=\gamma_+ m_f$ and $m^-_f=\gamma_-m_f$ which satisfy the diagonal constraints \eqref{quantum diag}. The vertex is a sum over the representations labelling faces of the new spin foam vertex $W_v$ \eqref{new vertex}, weighted by coefficients which depend on $m_f$: 
\beq
\tl{W}_v\bigl(m_f,k_{ft},l_t\bigr) = \sum_{J^\pm_f \geq \lv\gamma_\pm m_f\rv} W_v\bigl(J^+_f,J^-_f,k_{ft},l_t\bigr) \prod_f\Biggl[c^{J^+_f}_{\gamma_+m_f}\,c^{J^-_f}_{\gamma_-m_f}\,\prod_t \begin{pmatrix} J^+_f & J^-_f & k_{ft} \\ \gamma_+m_f & \gamma_-m_f & -(\gamma_++\gamma_-)m_f \end{pmatrix} \Biggr].
\ee
It is certainly not surprising that the vertex $W_v$ appears in this expression. Indeed, the main feature of $W_v$ is the fusion coefficient inserted on each tetrahedron. Such a coefficient is very natural with regards to the implementation of the cross-simplicity constraints: $b_{-ft}=-N_t\mone b_{+ft}N_t$. Since the latter relate the self-dual to the anti-self-dual part of $B_{ft}$, it is natural that some self-dual and anti-self-dual spins, or magnetic numbers, intertwine at the quantum level. To say it roughly, the main difference between the above vertex amplitude and the new models relies in the gluing of 4-simplices. The physical idea usually retained is that the area of a triangle has to be the same for all 4-simplices sharing it, since it is a gauge invariant quantity. Then, if area is encoded in representations, this naturally leads to the new FK and EPR models. However, the present work show that the information about areas may more generically be contained in some data which are rather angular momenta (magnetic numbers) than spin representations. From this perspective, the FK and EPR models are the simplest models, which remove the sums over representations $J^\pm_f$ in $\tl{W}_v$ by selecting the highest magnetic numbers through the choice: $c^j_m=\delta_{j,\lv m\rv}$.





\section*{Conclusion}

We have provided an action and a path integral on a triangulation of spacetime for the new spin foam models. This makes clear the link with geometry and Regge calculus. Indeed, the action consists in a Regge action, which includes the Immirzi parameter, and supplemented with the simplicity constraints and a non-trivial measure ensuring a consistent gluing of simplices. This measure is intimately related to the action of \cite{conrady1} and gives it an important geometric role, that of concentrating the amplitude around configurations which satisfy the correct rules of parallel transport. When these rules are satisfied, the dihedral angles entering the Regge action are functions of the 3d dihedral angles, as expected in area-angle Regge calculus \cite{area-angleRC}. It is then natural to embed the model within a class of models where the quantum weight of the parallel transport relations can be arbitrarily chosen. We have derived the corresponding spin foam models.

However, the closure relation does not generically hold for these models. The gluing function used in the FK and EPR models is a special case where it holds, so that on-shell configurations are Regge-like: they come from discretized co-tetrads. In this work, the quantum implementation of parallel transport is completely under control. It would be fine to similarly control the closure relation. The work of Conrady and Freidel \cite{conrady3} is certainly an important step in that direction. It gives in addition a beautiful description of both classical and quantum tetrahedra, and it may be useful to translate our results with their variables.

A related issue, due to the fact that the closure constraint is not {\it a priori} imposed, is that the equations of motion describe flat spacetimes. This feature should be displayed when summing over the triangle representations. It would be interesting to see that on precise situations, with given triangulations. Notice that it is crucial to know the effect of summing over triangle representations in coarse-graining processes. The situation is similar in computations of correlation functions for non-trivial triangulations \cite{eugenio}. We hope that the geometric content described in this work will help to answer these questions and to develop the techniques of \cite{graviton}.

Another interesting direction is to provide a group field formulation of the new models. The scheme proposed in \cite{daniele} makes the gluing explicit, in the spirit of the present paper. It is however necessary to introduce of the Immirzi parameter. We also wonder how the underlying Regge geometries can be unfold.

\section*{Acknowledgements}

The author thanks Etera R. Livine, Daniele Oriti and Simone Speziale for their interest in this work and their availibility to discuss spin foams.



\begin{thebibliography}{99}

\bibitem{1}
  V. Bonzom,
  {\it From lattice BF gauge theory to area-angle Regge calculus},
  arXiv:0903.0267 [gr-qc].

\bibitem{regge}
  T.~Regge,
  {\it General relativity without coordinates},
  Nuovo Cim.\  {\bf 19} (1961) 558.


\bibitem{barrett first order}
  J. W. Barrett,
  {\it First order Regge calculus},
  Class.\ Quant.\ Grav.\ {\bf 11} 2723 (1994)
  [arXiv:hep-th/9404124].


\bibitem{area-angleRC}
  B. Dittrich and S. Speziale,
  {\it Area-angle variables for general relativity},
  New J.\ Phys.\ {\bf 10} 083006 (2008)
  arXiv:0802.0864 [gr-qc].

\bibitem{baez def SF}
  J. C. Baez,
  {\it Spin foam models},
  Class. Quant. Grav. {\bf 15} 1827 (1998)
  [arXiv:gr-qc/9709052].

\bibitem{reisenberger worldsheet}
  M. P. Reisenberger,
  {\it Worldsheet formulations of gauge theories and gravity},
  [arXiv:gr-qc/9412035].

\bibitem{drouffe zuber}
  J.M. Drouffe and J.B. Zuber,
  {\it Strong coupling and mean field methods in Lattice Gauge Theories},
  Phys. Rept. {\bf 102} 1 (1983).


\bibitem{baez BF}
  J. C. ~Baez,
  {\it An introduction to spin foam models of quantum gravity and BF theory},
  Lect. Notes Phys. {\bf 543} (2000) 25
  [arXiv:gr-qc/9905087].

\bibitem{plebanski}
  J.~F.~Plebanski,
  {\it On the separation of Einsteinian substructures},
  J.\ Math.\ Phys.\  {\bf 18}, 2511 (1977).

\bibitem{simplicity}
  M.~P.~Reisenberger,
  {\it Classical Euclidean general relativity from *left-handed area = right-handed area*},
  [arXiv:gr-qc/9804061]. \\
  R. De Pietri and L. Freidel,
  {\it so(4) Plebanski action and relativistic spin foam model},
  Class. Quant. Grav. {\bf 16}, 2187 (1999)
  [arXiv:gr-qc/9804071].

\bibitem{epr}
  J.~Engle, R.~Pereira and C.~Rovelli,
  {\it Flipped spinfoam vertex and loop gravity},
  Nucl.\ Phys.\  B {\bf 798}, 251 (2008)
  arXiv:0708.1236 [gr-qc]. \\
  J.~Engle, E.~Livine, R.~Pereira and C.~Rovelli,
  {\it LQG vertex with finite Immirzi parameter},
  Nucl. Phys. B {\bf 799}, 136 (2008)
  arXiv:0711.0146 [gr-qc].

\bibitem{fk}
  L.~Freidel and K.~Krasnov,
  {\it A New Spin Foam Model for 4d Gravity},
  Class. Quant. Grav. {\bf 25}, 125018 (2008)
  arXiv:0708.1595 [gr-qc].

\bibitem{conrady1}
  F. Conrady and L. Freidel,
  {\it Path integral representation of spin foam models of 4d gravity},
  Class. Quant. Grav. {\bf 25}, 245010 (2008)
  arXiv:0806.4640 [gr-qc].


\bibitem{conrady2}
  F. Conrady and L. Freidel,
  {\it On the semiclassical limit of 4d spin foam models},
  Phys. Rev. D {\bf 78}, 104023 (2008)
  arXiv:0809.2280 [gr-qc].

\bibitem{conrady3}
  F. Conrady and L. Freidel,
  {\it Quantum geometry from phase space reduction},
  arXiv:0902.0351 [gr-qc].  

\bibitem{barrett fairbairn}
  J. W. Barrett, R. J. Dowdall, W. J. Fairbairn, H. Gomes and F. Hellmann,
  {\it Asymptotic analysis of the EPRL four-simplex amplitude},
  arXiv:0902.1170 [gr-qc].

\bibitem{quantum tet}
  A. Barbieri,
  {\it Quantum tetrahedra and simplicial spin networks},
  Nucl.\ Phys.\ B\ {\bf 51} (1998) 714-728,
  [arXiv:gr-qc/9707010]. \\
  J.C. Baez and J.W. Barrett,
  {\it The quantum tetrahedron in three-dimensions and four-dimensions},
  Adv. Theor. Math. Phys. {\bf 3} (1999) 815-850,
  [arXiv:gr-qc/9903060].

\bibitem{etera cs}
  E.~R.~Livine and S.~Speziale,
  {\it New spinfoam vertex for quantum gravity},
  Phys.\ Rev.\  D {\bf 76}, 084028  (2007)
  arXiv:0705.0674 [gr-qc].
  
\bibitem{etera new sf}
  E.~R.~Livine and S.~Speziale,
  {\it Solving the Simplicity Constraints for Spinfoam Quantum Gravity},
  Europhys.\ Lett.\  {\bf 81}, 50004  (2008)
  arXiv:0708.1915 [gr-qc].


\bibitem{lagrangian BC}
  V. Bonzom and E. R. Livine
  {\it A Lagrangian approach to the Barrett-Crane spin foam model},
  Phys. Rev. D {\bf 79}, 064034, (2009)
  arXiv:0812.3456 [gr-qc].

\bibitem{BC}
  J.W.~Barrett and L.~Crane,
  {\it Relativistic Spin Networks and Quantum Gravity},
  J.\ Math.\ Phys.\ {\bf 39}, 3296 (1998)
  [arXiv:gr-qc/9709028].


\bibitem{daniele}
  D. Oriti,
  {\it  Group field theory and simplicial quantum gravity},
  arXiv:0902.3903 [gr-qc].


\bibitem{caselle}
  M. Caselle, A. D'Adda and L. Magnea,
  {\it Regge Calculus As A Local Theory Of The Poincare Group},
  Phys. Lett. B {\bf 232} 457 (1989).

\bibitem{kawamoto}
  N. Kawamoto and H. B. Nielsen,
  {\it Lattice Gauge Gravity With Fermions},
  Phys. Rev. D {\bf43}, 1150 (1991).

\bibitem{abramo}
  M. Abramowitz and I. Stegun, editors.
  {\it Handbook of mathematical functions with formulas, graphs and mathematical tables},
  Dover Publications Inc., New York, 1992. Reprint of the 1972 edition.

\bibitem{ooguri}
  H.~Ooguri,
  {\it Topological lattice models in four-dimensions},
  Mod.\ Phys.\ Lett.\ A\ {\bf 7}, 2799 (1992)
  [arXiv:hep-th/9205090].

\bibitem{carlo 3d}
  C.\,Rovelli: 
  {\it The Basis of the Ponzano-Regge-Turaev-Viro-Ooguri quantum gravity model in the loop representation basis},
  Phys.\,Rev.\,{\bf D48}, 2702 (1993).


\bibitem{eugenio}
  E. Bianchi and A. Satz,
  {\it Semiclassical regime of Regge calculus and spin foams},
  Nucl. Phys. B {\bf 808}:546-568 (2009)
e-Print: arXiv:0808.1107 [gr-qc] 

\bibitem{graviton}
  E. R. Livine and S. Speziale,
  {\it Group integral techniques for the spinfoam graviton propagator},
  JHEP {\bf 0611}, 092 (2006),
  [arXiv:gr-qc/0608131].

\bibitem{carlo book}
  C.\,Rovelli,  \newblock {\em Quantum Gravity}
  \newblock (Cambridge University Press, Cambridge 2004).


\end{thebibliography}
\end{document}